\documentclass{ws-mpla}
\usepackage[super]{cite}
\usepackage{graphicx}
\usepackage{aas_macros}
\usepackage{hyperref}

\begin{document}

\markboth{Marcel S. Pawlowski}
{The Planes of Satellite Galaxies Problem}

\catchline{}{}{}{}{}

\title{THE PLANES OF SATELLITE GALAXIES PROBLEM,\\
SUGGESTED SOLUTIONS, AND OPEN QUESTIONS}

\author{\footnotesize MARCEL S. PAWLOWSKI\footnote{Hubble Fellow}}

\address{Department of Physics and Astronomy, University of California, \\
Irvine, CA 92697, USA\\
marcel.pawlowski@uci.edu}

\maketitle

\pub{Received (Day Month Year)}{Revised (Day Month Year)}

\begin{abstract}
Satellite galaxies of the Milky Way and of the Andromeda galaxy have been found to preferentially align in significantly flattened planes of satellite galaxies, and available velocity measurements are indicative of a preference of satellites in those structures to co-orbit. There is increasing evidence that such kinematically correlated satellite planes are also present around more distant hosts. Detailed comparisons show that similarly anisotropic phase-space distributions of sub-halos are exceedingly rare in cosmological simulations based on the $\Lambda$CDM paradigm. Analogs to the observed systems have frequencies of $\leq 0.5$\ per cent in such simulations. In contrast to other small-scale problems, the satellite planes issue is not strongly affected by baryonic processes because the distribution of sub-halos on scales of hundreds of kpc is dominated by gravitational effects. This makes the satellite planes one of the most serious small-scale problem for $\Lambda$CDM.
This review summarizes the observational evidence for planes of satellite galaxies in the Local Group and beyond, and provides an overview of how they compare to cosmological simulations. It also discusses scenarios which aim at explaining the coherence of satellite positions and orbits, and why they all are currently unable to satisfactorily resolve the issue.
\end{abstract}

\ccode{PACS Nos.: include PACS Nos.}

\section{Introduction}	

According to the $\Lambda$CDM model of cosmology, baryons constitute only $\approx 5\%$\ of the energy density of the Universe, which is dominated by dark energy (68\%, parametrized as a cosmological constant $\Lambda$) and cold dark matter (27\%, CDM).\cite{Planck2016} The $\Lambda$CDM model is currently widely accepted due to its successes in matching astrophysical observations on cosmic scales, such as the power spectrum of the Cosmic Microwave background, the large-scale structure traced by observed galaxies, the accelerated expansion of the Universe, and Big Bang Nucleosynthesis. However, successes do not constitute proof of correctness,\cite{McGaugh2015,Merritt2017} and science strives to probe the limitations of its models by pushing to test them in novel ways. In particular when expanded beyond its initial scope of applicability, a model is most thoroughly tested and opportunities for refinement or revision can be identified. For $\Lambda$CDM --- developed to fit the large-scale structure and evolution of the Universe (on scales of many Mpc to Gpc) --- such a novel domain of applicability is the regime of galaxies and their satellite galaxy systems (on scales of hundreds of kpc and below). Due to the non-linearity of $\Lambda$CDM on those scales, model predictions are made via numerical simulations. These evolve an initially almost homogenious matter distribution forward in time, simulating the formation and evolution of structure in the cosmos until the present day. This implies an important caveat for tests in this regime: not the model itself is tested directly, but its realisation via numerical simulations.

Our best observational knowledge of the smallest galaxies stems from observation of the $\approx 50$\ and 40 known dwarf Spheroidal (dSph) satellite galaxies of our Milky Way (MW) and the neighboring Andromeda galaxy (M31), respectively. Together with $\approx 15$\ more isolated dwarf galaxies, they make up the Local Group of galaxies, covering a volume of approximately 1\,Mpc in radius. Comparisons of cosmological simulations of dark matter halos resembling those expected to host the MW and M31 with the observed systems have revealed a number of small-scale problems of $\Lambda$CDM.\cite{Bullock2017} Among the best known issues are the Missing Satellites Problem, the Core-Cusp Problem, and the Too-Big-to-Fail problem. Yet, since the $\Lambda$CDM model is not tested directly but rather via simulations, these problems have largely not been seen as a failure of $\Lambda$CDM, but rather as teachable moments for galaxy formation theory. They can be addressed by incorporating baryonic feedback processes in the simulations, or by modifications to the cold dark matter particle (to warm or self-interacting dark matter). 

Yet, there is another, very basic property of satellite galaxy systems that can be utilized to test the $\Lambda$CDM model: their overall phase-space distribution; in other words the positions and velocities of satellite dwarf galaxies relative to their host galaxy and to each other. Especially for the highly dark matter dominated dSphs, their positions and systemic motions are largely independent of internal baryonic processes, strength of stellar feedback, and type of dark matter, but rather governed by the overall gravitational field dominated by the distribution of dark matter. It is also observationally easier to obtain the positions and systemic motion of dwarf galaxies than resolving their internal dynamics. This is because the latter might be affected by observational uncertainties of similar order than the to be resolved velocity dispersion (as low as $\approx 2\,\mathrm{km\,s}^{-1}$\ for the faintest systems), contributions from binary star motions, and non-equilibrium dynamics. The phase-space distribution of satellite galaxies thus can act as a rather robust test of a cosmological model, and the dark matter hypothesis.

Observations indicate that a substantial fraction of satellite galaxies are part of highly flattened, planar arrangements. Available kinematic information furthermore supports the notion that these structures might be rotating. This is in stark constrast to the much more random distribution and motion exhibited by satellite sub-halos around their host galaxy halo in cosmological simulations. This mismatch between the observed, spatially and kinematically coherent planes of satellite galaxies and $\Lambda$CDM cosmological simulations which contain only a very low frequency of satellite systems with comparable arrangements, has become known as the {\it Planes of Satellite Galaxies Problem}.


\section{Observed Planes of Satellite Galaxies}
\label{sect:observed}

\begin{figure}[ph]
\centerline{\includegraphics[width=2.65in]{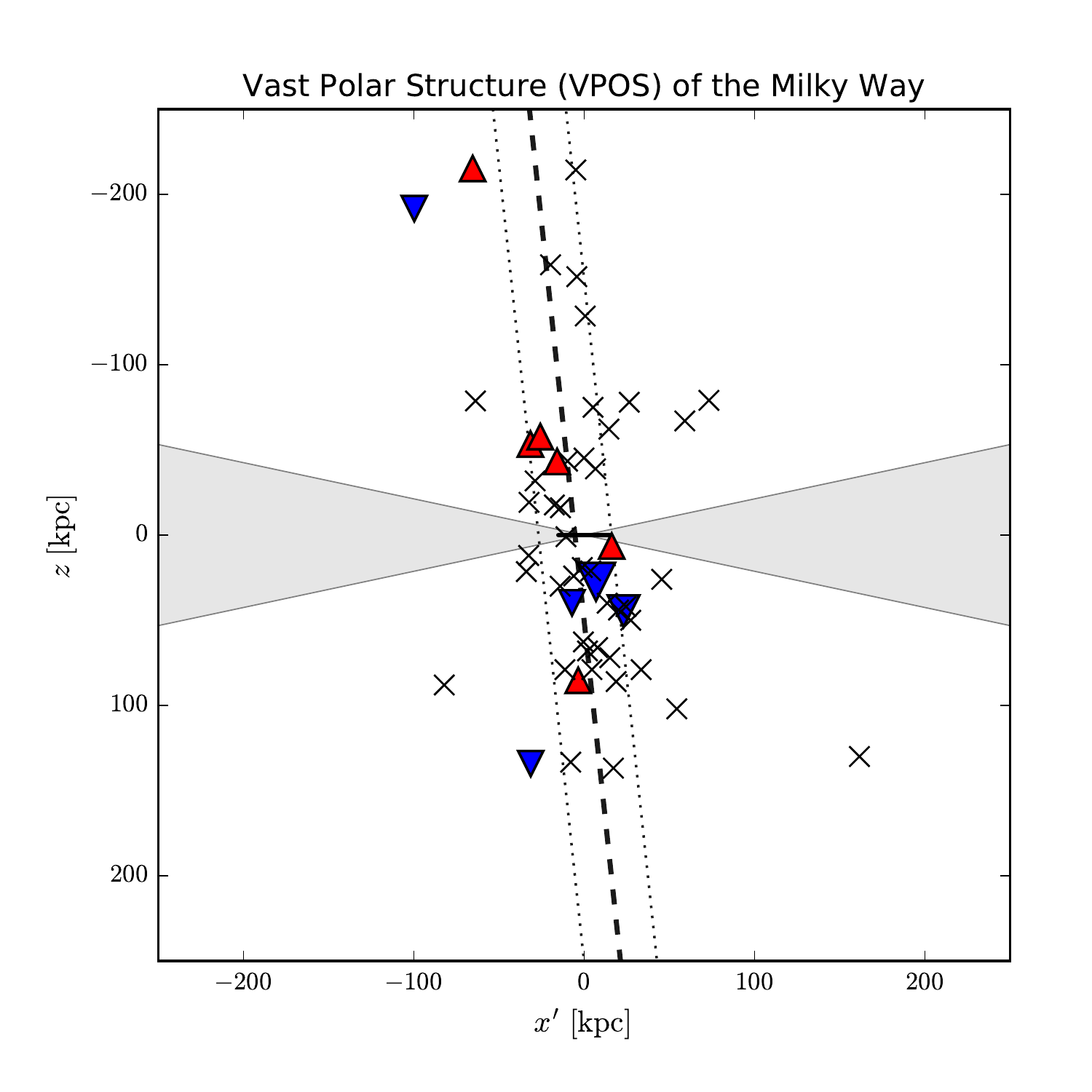}\includegraphics[width=2.65in]{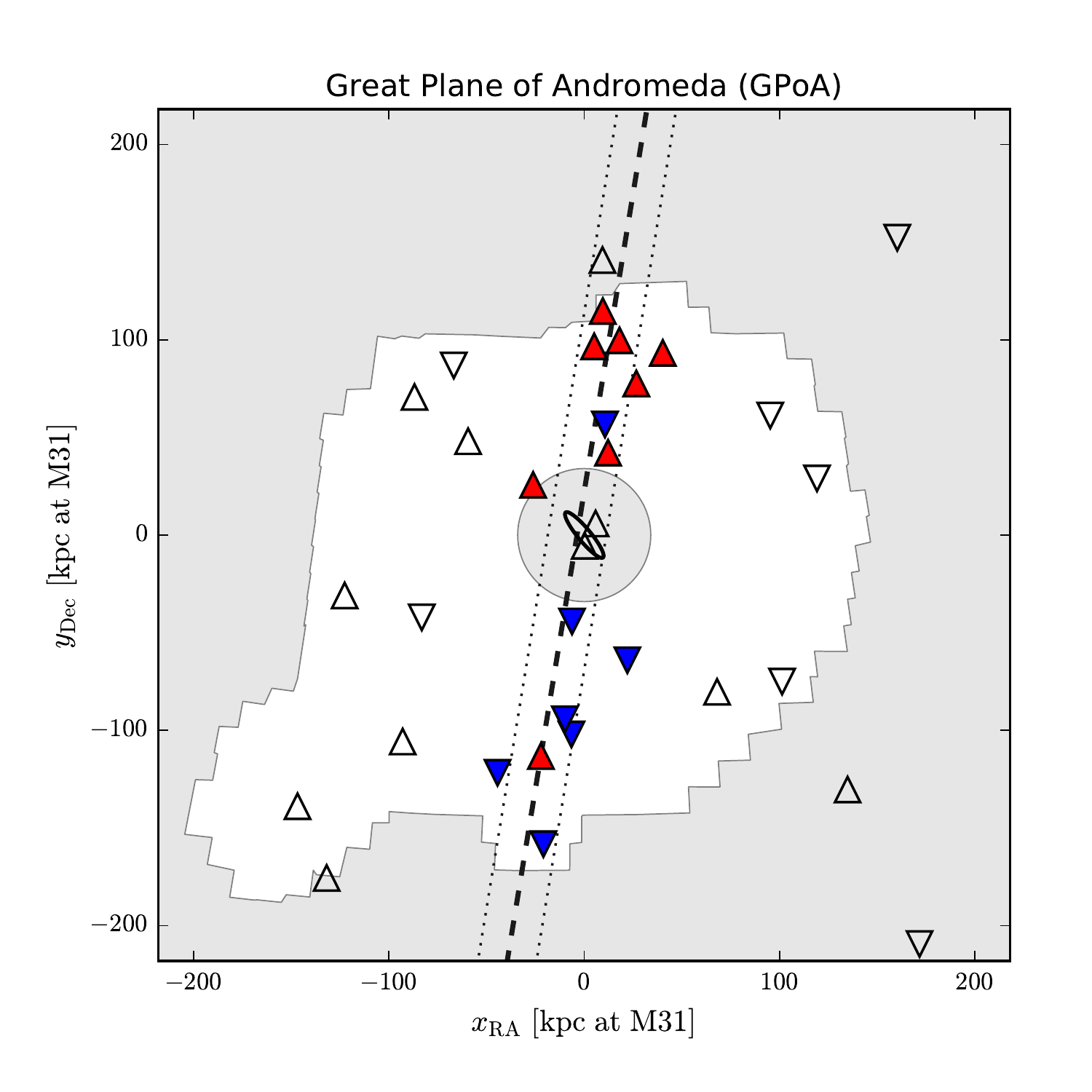}}
\centerline{\includegraphics[width=2.65in]{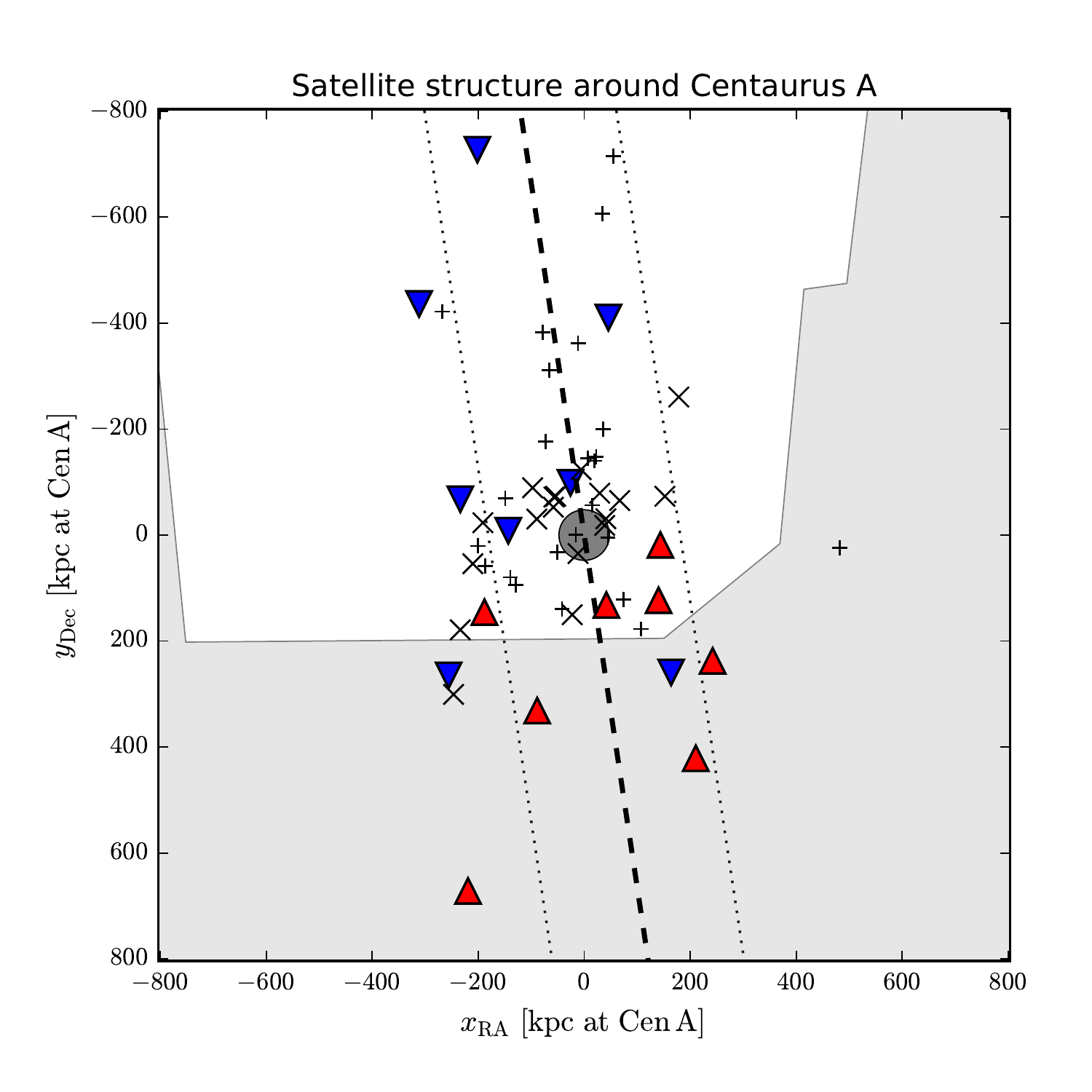}}
\vspace*{8pt}
\caption{
The three most prominent known planes of satellite galaxies in approximately edge-on orientations. Panel 1: the VPOS of the MW as seen from a position in which both the Galaxy (black line) and the satellite plane are seen edge-on. Panel 2: the GPoA around M31 (black ellipse) as seen from the Sun. Panel 3: the satellite structure around Centaurus\,A (grey circle) as seen from the Sun. The orientations and widths of the best-fit satellite planes in these views are indicated with dashed and dotted lines, respectively. Satellite galaxies with measured velocities are color-coded according to whether they are approaching (blue, downward triangles) or receding (red, upward triangles) relative to their host in these views, indicating coherent kinematics consistent with satellites co-orbiting in the planes. Satellites for which no proper motions (MW) or line-of-sight velocities (Cen\,A) are known are plotted as crosses, M31 satellites that are not part of the GPoA are plotted as open triangles whose orientation indicates the line-of-sight velocity direction. The grey areas denote regions with substantial observational limitations: the region $\pm 12^\circ$\ from the MW disk plane that is considered obscured by Galactic foreground\cite{Pawlowski2016} (panel 1), the region outside of the PAndAS survey around M31\cite{McConnachie2009} (panel 2), and the region outside of the DEC survey around Centaurus\,A\cite{Mueller2017a} (panel 3). 
 \protect\label{fig1}
 }
\end{figure}

This section provides an overview of the observational evidence for planes of satellite galaxies in the Local Group (Sects. \ref{sect:VPOS}, \ref{sect:GPoA}, and \ref{sect:LGplanes}) and beyond (Sect. \ref{sect:beyondLG}). These planes are typically characterized by their small root mean square (rms) height $r_{\perp}$\ measured in absolute distance, or by their low minor-to-major axis ratio $c/a$ measuring their relative rms flattening along these axes. 
Typically, the significance of a satellite plane is given for a null hypothesis of an isotropic satellite distribution, while accounting for survey footprints and obscured regions as applicable. This is not to be confused with the frequency of analog systems in cosmological simulations, because simulated satellite systems are mildly anisotropic (see Sect. \ref{sect:theproblem} for a discussion of comparisons to simulations). 
A detailed comparison of the parameters and mutual orientation of the satellite and dwarf galaxy planes in the Local Group can be found in Ref.~\refcite{Pawlowski2013a}. Figure \ref{fig1} shows the three most prominent observed planes of satellite galaxies.

\subsection{The Vast Polar Structure (VPOS) of the Milky Way}
\label{sect:VPOS}

First indication for an anisotropic distribution of MW satellite galaxies were discussed more than 40 years ago, when the then-known satellite galaxies -- the Large and Small Magellanic Clouds (LMC, SMC), Draco, Ursa Minor, Sculptor, Fornax, Leo\,I and II -- and globular clusters (GCs) at distances beyond 25\,kpc were found to lie along the same polar great circle as the Magellanic Stream.\cite{LyndenBell1976,Kunkel1976} This was followed over the next decades by several studies that compared the distributions of dSphs and GCs around the MW and proposed several streams of associated objects.\cite{Lynden-Bell1982,Majewski1994,Lynden-Bell1995,Pecci1995,Palma2002}

It was Ref.~\refcite{Kroupa2005} who first compared the MW satellite galaxy system to cosmological expectations in order to test the $\Lambda$CDM model. They investigated the positions of the 11 then-known ``classical'' satellite galaxies ($M_\mathrm{V} < -8$), and found them to align in a narrow plane that is oriented almost perpendicular to the disk of the MW. They argued that the observed satellite distribution is very rare if the satellite positions were drawn from an isotropic distribution, and reject this hypothesis with 99.6 per cent confidence if considering the positions of all 11 classical satellite galaxies.

Today, we know that the Milky Way is surrounded by a {\it Vast Polar Structure (VPOS, top left panel in Fig. \ref{fig1})}, a polar, flattened distribution of satellite galaxies and distant globular clusters.\cite{Pawlowski2012a} Depending on the exact sample of considered satellite galaxies, their distribution is fit by a plane with rms height $r_{\perp} = 20~\mathrm{to}~30$\,kpc, an axis ratio $c/a$\ of 0.18 to 0.30, and an inclination of 73$^\circ$\ to 87$^\circ$\ relative to the MW plane.\cite{Metz2007,Pawlowski2015a} In addition to satellite galaxies and globular clusters, up to half of the known stellar and gaseous streams in the MW halo beyond 10\,kpc align to within $35^\circ$\ or better with the plane defined by the satellite galaxies.\cite{Pawlowski2012a} The Magellanic Stream, the most prominent stream in the MW halo, is among the best-aligned streams. Since streams are debris shed along the orbits of tidally disrupting satellite galaxies or star clusters, this alignment indicates a preference for satellites to orbit along the VPOS.

A more direct access to the orbital directions of satellite galaxies is provided by proper motion measurements, which have provided tangential velocities for the 11 classical MW satellite galaxies. Of these, at least 8 are consistent with orbital planes aligned closely with the VPOS.\cite{Metz2008,Pawlowski2013b,Pawlowski2017a} The satellite galaxies Carina, Draco, Fornax, Leo\,II, LMC, SMC, and Ursa Minor all appear to co-orbit along the VPOS, while Sculptor is counter-orbiting in the structure. Such a tight alignment of orbital directions is very rare. When accounting for the obscuration by the disk of the MW, the spatial flattening of the 11 classical satellite galaxies alone has a significance of $2.5\,\sigma$\ relative to an isotropic null hypothesis. Combined with the tight orbital alignment, usually expressed as the concentration of orbital poles (directions of orbital angular momentum around the host), this rises to $4\,\sigma$.\cite{Pawlowski2016}

Discoveries of additional, fainter satellite galaxies, predominantly via the Sloan Digital Sky Survey (SDSS)\cite{York2000}, and the Dark Energy Survey (DES)\cite{DES2005}, have further supported this preferred spatial alignment, Most of the newly discovered satellite galaxies were found to align well with the plane defined by the more luminous MW satellites.\cite{Metz2009b,PawlowskiKroupa2014,Pawlowski2015a} Of concern in this regard is that the distribution of these discoveries is affected by the uneven sky coverage of such surveys. In particular for the SDSS, which has a footprint aimed at the Galactic caps, an agreement of additional objects with a polar structure might be expected irrespective of the underlying distribution of satellites. By comparing with mock-observations of well defined satellite distributions, this effect was found to be insufficient to explain the strong agreement of the fainter MW satellite galaxies with the plane defined by the 11 classical satellites, thereby demonstrating that the additional satellites indeed increase the significance of the VPOS, to $5\,\sigma$\ if combined with the spatial and kinematic correlation of the classical satellite galaxies.\cite{Pawlowski2016}

\subsection{The Great Plane of Andromeda (GPoA)}
\label{sect:GPoA}

Early studies had found that the overall M31 satellite system is not strongly flattened, but had seen hints that subsamples of satellites align in common planes.\cite{Koch2006,McConnachieIrwin2006,Metz2007} These studies suffered from a small sample of satellite galaxies and inhomogenious distance measurements, which complicated a full statistical analysis. This was alleviated by the PAndAS survey,\cite{McConnachie2009} which discovered additonal M31 satellite galaxies and provided homogeneous distance measurements for all known M31 satellite galaxies within the well-defined survey volume.\cite{Conn2012}
Using this dataset, Refs. \refcite{Ibata2013} and \refcite{Conn2013} identified a significant spatial satellite galaxy plane (0.13 per cent chance to occur in random satellite samples), consisting of a subset of 15 out of the 27 satellite galaxies in the survey volume. This {\it Great Plane of Andromeda (GPoA, top right panel in Fig. \ref{fig1})} is oriented almost edge-on as seen from the Sun, has a rms plane height of $r_{\perp} \approx 12.6$\,kpc, extends over 400\,kpc, and has an axis ratio of $c/a \approx 0.1$\cite{Pawlowski2013a}. What makes this structure more intriguing is that the line-of-sight velocities of the on-plane satellites show a strong correlation. Relative to M31, satellites in the north of M31 recede from us, while those in the south approach us. Of the 15 on-plane satellite, 13 follow this kinematic trend, which increases the structure's significance to 99.998 per cent. Due to the near-edge-on orientation, this kinematic correlation is consistent with a rotating plane of satellite galaxies, and reminiscent of the co-orbiting nature of MW satellites along the VPOS. If interpreted as a rotating plane of satellites, it implies a similar spin direction as for the VPOS.\cite{Pawlowski2013a}  However, it will require precise measurements of proper motions to determine whether the on-plane satellites do or do not have motions perpendicular to the GPoA.

Regarding their internal properties (sizes, luminosities, masses, metallicities, star formation histories), the sample of on-plane satellite galaxies of M31 does not differ from that of the off-plane satellites.\cite{Collins2015} However, spatially the system of on-plane satellite galaxies is strongly lopsided, with 13 of the 15 objects on the MW side of M31, whereas the off-plane satellites are more evenly distributed.\cite{McConnachieIrwin2006,Conn2013} The GPoA is also aligned with the Giant Stellar Stream in the M31 halo.\cite{Conn2013} In contrast to the VPOS, the GPoA is not perpendicular to the galactic disc of M31, but inclined by $\approx 50^\circ$. Despite some studies stating that the VPOS and GPoA are aproximately perpendicular,\cite{Conn2013} the two planes are in fact inclined by only 40 to 50$^\circ$\cite{Pawlowski2013a}.

\subsection{Non-satellite dwarf galaxy planes in the Local Group}
\label{sect:LGplanes}

The existence of planes of satellite galaxies around both major Local Group galaxies, both sharing similar spatial and kinematic correlations, with one being oriented towards the other, and both spinning in the same general direction, naturally raises the question whether there is a connection between them and the Local Group. Indeed, almost all of the $\approx 15$\ dwarf galaxies in the Local Group that lie beyond both the virial volumes of the MW and M31 are contained in one of two planar structures.\cite{Pawlowski2013a,PawlowskiMcGaugh2014} These {\it Local Group Planes (LGP1, LGP2)} are highly symmetric (both are offset from the MW and M31 by $\approx 300$\,kpc), strongly flattened (rms heights of $r_{\perp} \approx 60$\,kpc, rms long-to-short axis ratios of $c/a \approx 0.1$), and extend over 1 -- 2 Mpc. Without considering possible kinematic correlations, finding two spatial planes of similar or more extreme axis-ratio flattening occurs in only 0.3 per cent of random realisations.\cite{PawlowskiMcGaugh2014} The more prominent LGP1 extends between the MW and M31 along the direction of the Magellanic Stream, with its member dwarf galaxies following a similar line-of-sight velocity trend as the gas belonging to the stream. Such alignments might provide importent clues for possible formation scenarios of the satellite planes.

\subsection{Satellite structures beyond the Local Group}
\label{sect:beyondLG}

Searching for planes of satellite galaxies beyond the Local Group is challenging: faint satellite galaxies are more difficult to discover due to their low surface brightness, uncertainties in distance measurements based on the tip of the red giant branch method are of the order of 5 to 10 per cent and thus encompass the $\approx 250$\,kpc virial radius of a MW-like galaxy at a distance of $\gtrsim 5$\,Mpc, making a full three-dimensional spatial analysis impossible. Spectroscopic follow-up to confirm membership of a satellite candidate to a host galaxy and for investigations of kinematic correlations demand substantial observational ressources. Nevertheless, some indications of additional correlated satellite systems have been reported. These include a flattening of the dSph satellite galaxies around M81, \cite{Chiboucas2013} 
a narrow dwarf galaxy structure around M101,\cite{Mueller2017b} as well as numerous alignments of 2 or 3 dwarf galaxies with stellar streams (see Table 2 in Ref.~\refcite{PawlowskiKroupa2014}). Two particular approaches warrant further discussion: a detailed study of the satellite galaxy system around Centaurus\,A, and the first attempt at statistical investigations based on a larger sample of satellite systems identified in the SDSS.

\subsubsection{The Centaurus A Satellite Plane (CASP)}
\label{sect:CenA}

Centaurus A is a radio-active elliptical galaxy at a distance of about 3.8\,Mpc. Ref.~\refcite{Tully2015} have reported that of the 29 satellite galaxies with good distance measurements, all but two can be assigned to two parallel planes of satellites. These planes are seen almost edge-on from the MW, have rms heights of $r_{\perp} \approx 60$\,kpc, long-axis dimensions of $\approx 300$\,kpc (axis ratios $c/a \approx 0.2$), and are seperated by about 300\,kpc. The structure is highly significant, two similarly narrow planes are found in only 0.03 per cent of random realisations that maintain the radial distribution and overall flattening of the Centaurus A satellite system. \cite{Tully2015}

An analysis including 29 additional, more recently discovered satellite galaxies and candidates\cite{Crnojevic2014,Crnojevic2016,Mueller2017a} has since challenged the double-planar nature of the satellite system, but finds evidence for the existence of one, similarly oriented but wider, plane of satellite galaxies.\cite{Mueller2016} Interestingly, this satellite plane is approximately perpendicular to the dusty plane of Centaurus A, similar to the polar orientation of the VPOS.\cite{Mueller2016}
A more precise determination of the satellite plane properties and a final verdict on the double- or single-planar structure will require accurate distance measurements to the satellite candidates, to confirm their membership and map the full three-dimensional structure of the system.

Line-of-sight velocity measurements are available for 16 of the confirmed Centaurus A satellite galaxies. They reveal a kinematic correlation akin to that observed for the GPoA: Satellites in the north of Centaurus A approach relative to the host, while those in the south recede.\cite{Mueller2018} {\it The Centaurus A Satellite Plane (CASP, bottom panel in Fig. \ref{fig1})} thus appears to rotate. Of the 16 satellites with velocity measurements, 14 follow this common trend (a similar signal occurs in only 0.4 per cent of random cases, the significance for the kinematics alone is $2.6\,\sigma$). The kinematic trend is aligned with the orientation of the satellite planes. This joint spatial and kinematic correlation can be considered strong evidence for a satellite plane, and it emphasises the similarity to the VPOS and GPoA and their kinematic signatures consistent with rotating satellite planes. The Centaurus A system is the best-studied case for a plane of satellites beyond the Local Group thus far. It indicates that the occurence of planes of satellite galaxies is not confined to the Local Group, nor to satellite systems around spiral galaxies of MW and M31 mass (virial mass $M_{\mathrm{vir}} \approx 1~\mathrm{to}~3~\times~10^{12}\,M_{\odot}$), but can also be found around a more massive elliptical galaxy ($M_{\mathrm{vir}} \approx 10^{13}\,M_{\odot}$).

\subsubsection{Statistical approaches}
\label{sect:statistical}

The difficulty in studying planes of satellite galaxies around more distant host galaxies lies in the need to measure mutual orientations and motions of satellite galaxies, while even under ideal conditions only projected positions and line-of-sight velocities are observationally accessible. Furthermore, for more distant host galaxies only a very of the few brightest satellites are known, making it more difficult to compare to the Local Group satellite planes.

A pioneering study aimed at statistically measuring the incidence of planes of satellite galaxies attempted to overcome these limitations by selecting satellites on diametrically opposite sides of their host galaxies.\cite{Ibata2014} They selected systems with at least two satellite galaxies within a projected seperation of 150\,kpc around M31-like hosts at redshift $z < 0.05$, and for which spectroscopic velocities are available for both satellites. Requiring the satellites to be on opposite sides to within a small opening angle maximizes the chance to see potential satellite planes edge-on. This in turn allows to use line-of-sight velocities to search for a signal of co-orbiting planes. If most satellites co-orbit in narrow planes, then the velocities should be anti-correlated relative to the host (one satellite approaches us while the other recedes in the host's rest frame). If the satellites are predominantly unrelated, half the sample should show the same velocity sign relative to the host, and only half should be anti-correlated. Out of 22 pairs of satellites identified in SDSS that lie on opposite sides to within an opening angle of $\leq 8^\circ$, 20 show a velocity anti-correlation (significance $> 4\,\sigma$). This excess of anti-correlated satellite velocities has been interpreted as being consistent with $> 50$\ per cent of the satellites around the hosts living in co-orbiting planes of satellites. \cite{Ibata2014} This interpretation is supported by an independend analysis which found an excess of satellite galaxy candidates positioned diametrically opposite to a spectroscopically confirmed satellite galaxy around hosts selected from the SDSS.\cite{Ibata2015} An equivalent signal is not seen in $\Lambda$CDM simulations.

While the velocity anti-correlation signal itself was confirmed by other studies, its interpretation as evidence for rotating planes of satellites has been challenged.\cite{Phillips2015,Cautun2015a} Two main concerns are that (1) the signal is less strong if satellite galaxies at larger radii from the hosts are included in the analysis, and that (2) no overabundance of velocity anti-correlation is found for larger opening angles. Both criticisms deviate from the originally intended analogy with the GPoA system. A reduced signal for satellite galaxies at larger projected distances from their host might merely hint at the typical size of the correlated satellite structures,\cite{Ibata2015} and testing what fraction of hosts have all their satellite galaxies exclusively distributed in a co-orbiting plane is not equivalent with the original measure which estimated what fraction of all satellites is part of co-orbiting planes. This is in line with the finding that for both the MW and M31 up to half of the satellite galaxies are not part of the planar structures\cite{Ibata2013,Conn2013,Pawlowski2016}. Ultimately, however, there is general agreement that it will require a larger set of satellite systems with measured line-of-sight velocities to settle this debate.

\section{The problem: comparison with $\Lambda$CDM simulations}
\label{sect:theproblem}

\begin{figure}[ph]
\centerline{\includegraphics[width=2.65in]{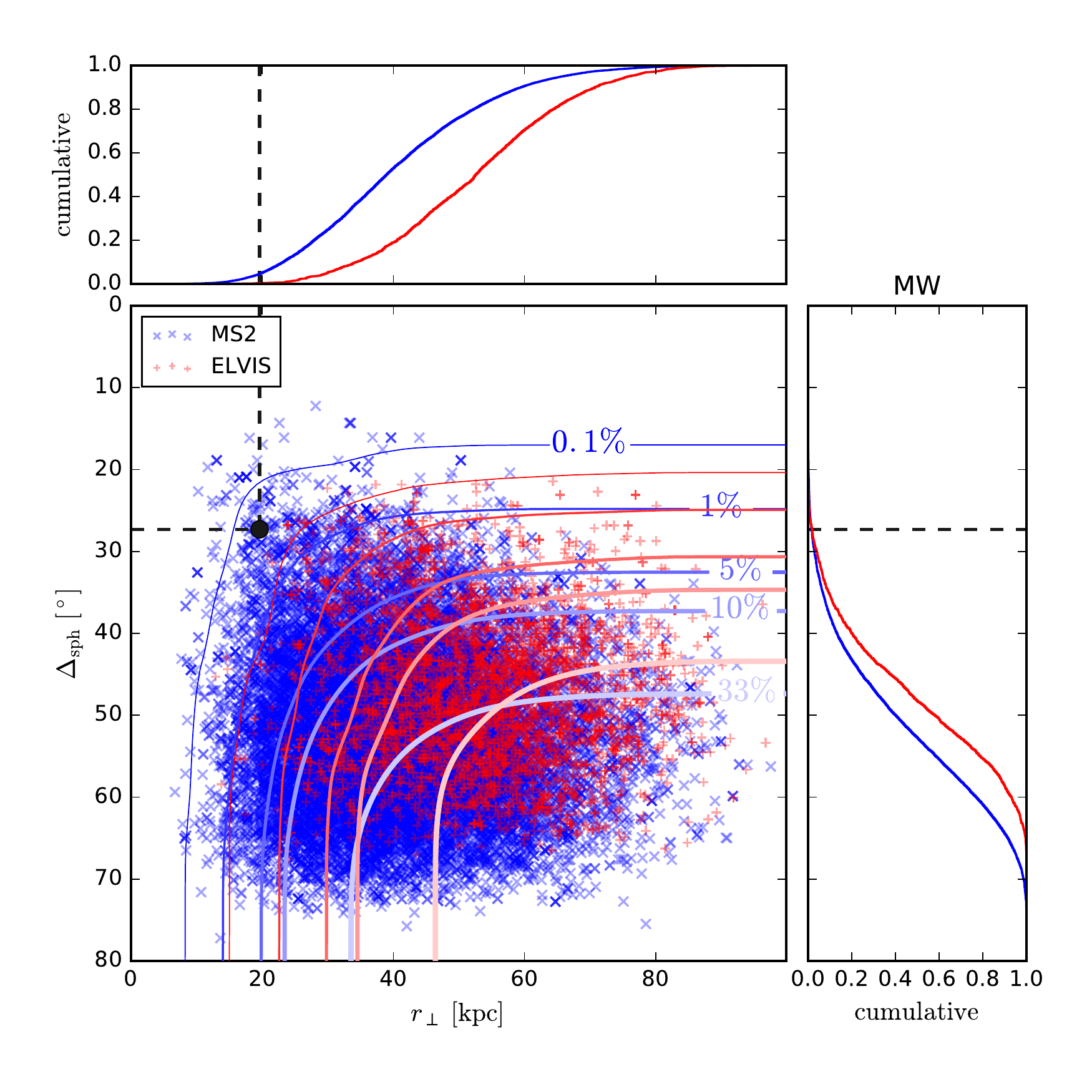} \includegraphics[width=2.65in]{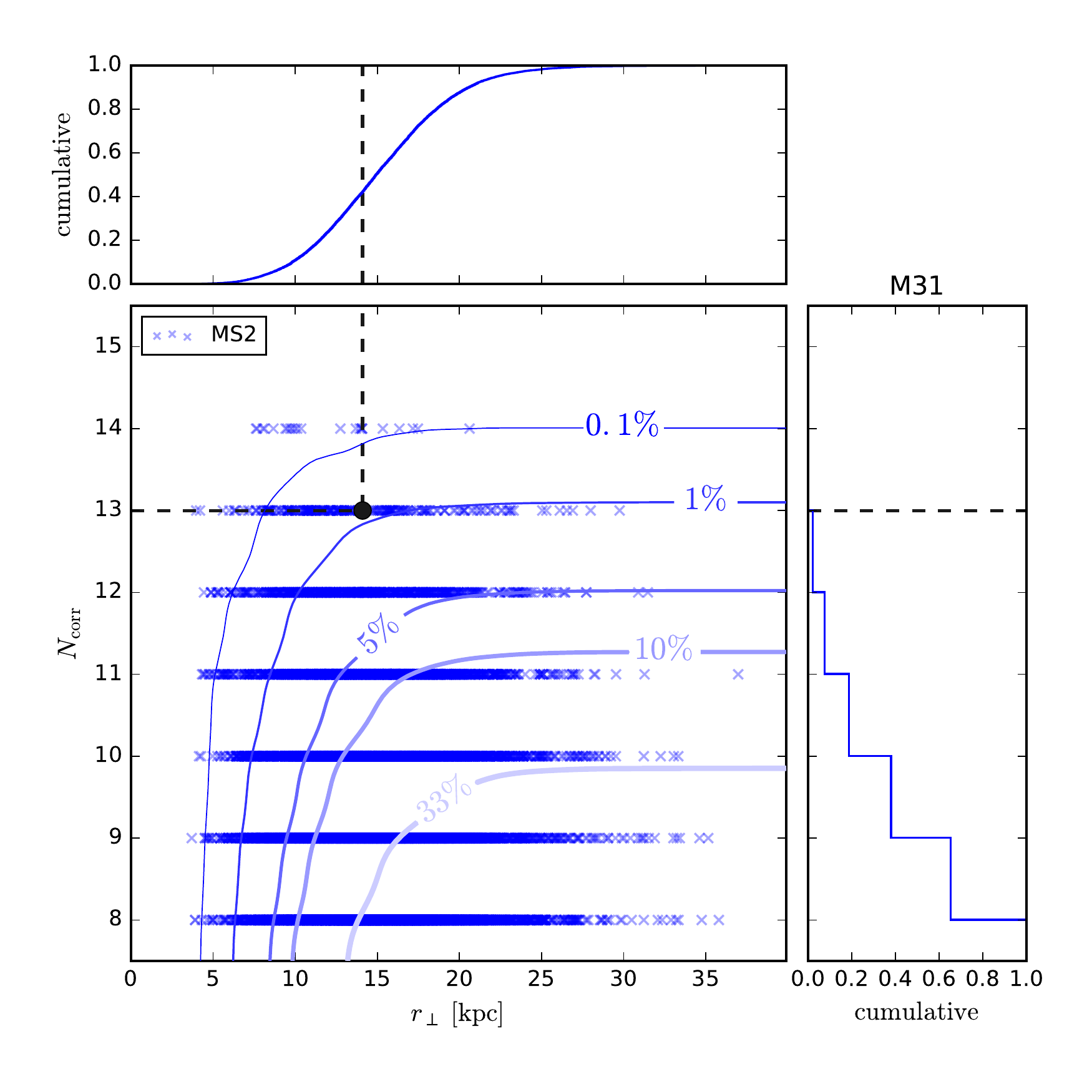}} 
\centerline{\includegraphics[width=2.65in]{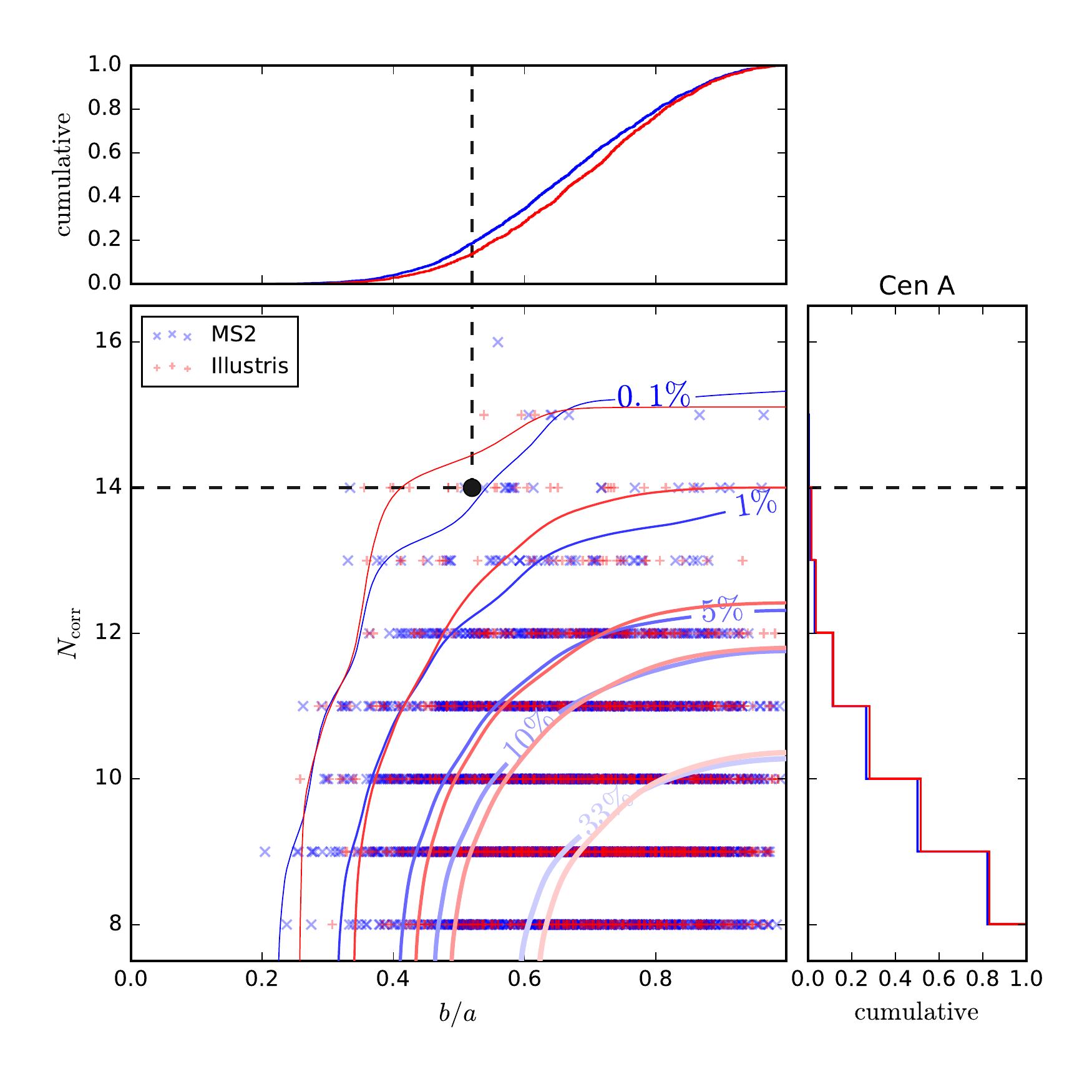}}
\caption{
Comparison of the three most pronounced observed satellite galaxy planes with cosmological simulations. Left: the VPOS of the MW (considering only the 11 classical satellites) compared to the Millennium-II and ELVIS simulations.\cite{Pawlowski2014,PawlowskiMcGaugh2014b} Right: the GPoA around M31 consisting of 15 out of 27 satellite galaxies found within the PAndAS survey footprint, compared to equivalent sub-sample satellite planes in the Millennium-II simulation.\cite{Pawlowski2014} Bottom: the CASP around Centaurus\,A compared to the dark-matter-only Millennium-II as well as the hydrodynamic Illustris simulation.\cite{Mueller2018} The horizontal axes give a measure of the plane flattening (rms height $r_\perp$\ for the MW and M31, on-sky short-to-long axis ratio $b/a$\ for Centaurus\,A), and the vertical axes give a measure of kinematic coherence (the spherical standard distance $\Delta_\mathrm{sph}$\ of the eight most-concentrated orbital poles for the MW, and the number of satellites $N_\mathrm{corr}$\ following the same line-of-sight velocity trend for M31 and Centaurus\,A). The side panels give the corresponding cumulative distribution in the flattening (top) and coherence (right) quantities. 
Each cross corresponds to one realisation of a simulated satellite system. The flattening and coherence of the observed structures are illustrated with dark dashed lines. The contour lines illustrate the frequency of at least as extreme satellite systems (i.e. as or more narrow {\it and} as or more kinematically coherent) as the observed system. That the observed planes of satellite galaxies tend to be 0.1\,\% outliers is the heart of the Planes of Satellite Galaxies Problem.
 \protect\label{fig2} }
\end{figure}

For the $\Lambda$CDM model to be a successful description of the Universe, it has to explain the existence of the observed planes of satellite galaxies. Satellite planes were not predicted by $\Lambda$CDM. Structure formation and the assembly of increasingly massive halos by accretion and merger events result in quite chaotic distributions of satellite sub-halos. Nevertheless, due to the preferential accretion of sub-halos along filaments of the cosmic web, the tendency of sub-halos to be accreted not individually but in small groups, and the overall triaxiality of the dark matter host halos, some anisotropy in the distribution of satellite galaxies is expected in $\Lambda$CDM. Thus, a priori it is unclear whether arrangements comparable to the observed planes of satellite galaxies are sufficiently frequent in cosmological simulations to be considered common and unremarkable, or so infrequent that they pose a problem for the model. It is therefore necessary to test $\Lambda$CDM by searching for structures similar to the observed satellite galaxy planes in simulations.

It should be noted that all comparisons between the observed planes of satellites and simulated satellite systems are thus far done without accounting for the effects of observational uncertainties, such as in distance or proper motion measurements. While these affect the observed satellite galaxy distribution, exact positions and velocities are used for simulated satellite galaxy positions and velocities. Because uncertainties tend to wash out tight correlations -- distance uncertainties result in wider measurements of plane heights than the intrinsic one, proper motion uncertainties disperse highly clustered orbital poles -- any stated frequency of finding satellite planes in simulations that are as extreme as the observed ones must be considered as only an upper limit.

Ref.~\refcite{Kroupa2005} were the first to identify the observed, planar distribution of satellite galaxies (they termed it the Disk of Satellites) as a problem for $\Lambda$CDM. They argued that the observed structure of 11 then-known classical satellite galaxies is inconsistent with being drawn from an isotropic distribution -- which they assumed to be a first-order approximation to the distribution of sub-halo satellites in cosmological simulations based on then-published $\Lambda$CDM results -- at 99.6 per cent confidence. This publication initiated a vigorous and at times controversial debate about the severity of the tension between the observed satellite galaxy plane and cosmological simulations.

Subsequently, it was pointed out that the sub-halo distribution in cosmological simulations is not isotropic.\cite{Zentner2005,Libeskind2005} This is because of the preferential accretion of sub-halos along cosmic filaments, and the intrinsic triaxiality of dark matter halos (see Sect. \ref{sect:filaments}). However, while satellite systems in $\Lambda$CDM differ significantly from isotropy, this is not identical with their anisotropy being sufficient to account for the highly correlated observed planes of satellite galaxies. Consequently, detailed comparisons to satellite systems in $\Lambda$CDM simulations are required in order to capture the intricacies of structure formation and satellite accretion in a fully cosmological context. These showed that it is indeed possible to find similarly flattened spatial distributions, as measured via the $c/a$ axis ratio, for the 11 classical MW satellite galaxies with a frequency of 4 to 6 per cent in cosmological simulations.\cite{Wang2013,Pawlowski2014}

However, finding true analogs of the observed situation requires reproducing the full phase-space structure, in particular the orbital alignment of a majority of the 11 classical MW satellite galaxies with their spatial orientation.\cite{Metz2008,Pawlowski2013b} Ref.~\refcite{Libeskind2009} was one of the first to compare the observed alignment of orbital planes of the MW satellites with their spatial structure with cosmological simulations. At the time, three of the 11 classical satellite galaxies had orbital poles aligned to better than $30^\circ$, which occurred in about 20 per cent of the considered simulations and was thus deemed not in tension with $\Lambda$CDM. This number is in agreement with other studies.\cite{Deason2011} However, additional and more accurate proper motion measurements have since increased the number of well aligned orbits to at least six and possibly eight,\cite{Pawlowski2013b} which occurs in $\lesssim 1$\ per cent of those simulations (without simultaneously requiring them to also match the spatial flattening). This is further supported by a comparison to the Millennium-II simulation, which finds that the orbital alignment is reproduced in 1 to 2.5 per cent of analog systems. The flattening as well as orbital alignment are simultaneously reproduced in only $\approx 0.3$\ per cent of all simulated systems for 11 satellites.\cite{Pawlowski2014} Figure \ref{fig2} shows a comparison of the observed flattening and orbital coherence of the 11 classical satellites with satellite systems in cosmological simulations in the top left panel, and similar comparisons for the GPoA and CASP in the top right and bottom panels, respectively.

One concern in using the Millennium-II simulation is its relatively low resolution (dark matter particle mass of $8.5 \times 10^6 \mathrm{M}_\odot$). To overcome the resulting issue of only barely resolved satellite galaxies, comparisons have also been performed for higher-resolution zoom simulations. One particularly interesting set of simulations is the Exploring the Local Volume in Simulations (ELVIS) suite\cite{GarrisonKimmel2014}. ELVIS is  a set of 12 host halo pairs selected to resemble the Local Group in mass, extent, and relative velocity of the MW and M31, as well as a comparison sample of 24 isolated host halos of comparable mass. Comparing with the VPOS, specifically with the flattening and orbital coherence of the 11 classical MW satellite galaxies, has confirmed the tension found for the Millennium-II simulation.\cite{PawlowskiMcGaugh2014b} Only one out of 4800 (0.2 per cent) realisation of mock-observed analogs to the system of 11 classical MW satellite galaxies are found to be both as flattened and as closely co-orbiting. This is below the frequency found in lower-resolution simulations (see also Fig. \ref{fig2}), indicating that increased resolution does not help to resolve the Planes of Satellite Galaxies Problem.

The discovery of the GPoA around M31 has further fueled the debate about the Planes of Satellite Galaxies Problem. Analogs of the GPoA have been searched in the Millennium-II simulation by requiring simulated systems to contain satellite planes of the same number of satellites as observed, which simultaneously display at least the observed flattening, extent, and degree of kinematic coherence. This resulted in a frequency of only 0.04 to 0.17 per cent,\cite{Ibata2014a,Pawlowski2014}.
Other authors have claimed less severe tension with the same $\Lambda$CDM simulation and reported that 2 per cent of satellite systems in this simulation reproduce the observed spatial flattening and kinematic correlation.\cite{Bahl2014} However, their analysis was critisized for not requiring simulated satellite systems to simulataneously reproduce the kinematic correlation and both structural parameters of satellite planes (radial extend and thickness), and for their substantially different selection of model satellites compared to the selection of observed satellites in number, selection area and distance from M31.\cite{Ibata2014a,Pawlowski2014} This resulted in an over-estimation of the frequency of similar structures in $\Lambda$CDM.

Another approach motivated by the GPoA, which consists of only 15 out of the 27 satellite galaxies in PAndAS, is to account for the look-elsewhere effect.\cite{Cautun2015b} This intends to consider the possibility of having found a different number of satellite galaxies in an as significant arrangement as the observed one (e.g. only 12 satellites, but in a more narrow distribution). The authors argue that not the same distribution as the observed GPoA should be counted in cosmological simulations, but distributions that are as significant in relation to an isotropic null hypothesis. This results in a frequency of halos with at least as prominent ``planes'' as the observed ones of 9 per cent for the GPoA, and 5 per cent for the VPOS (considering only the 11 classical satellites).
However, since both satellite systems in the Local Group display planes of satellite galaxies, the frequencies can be combined to give the frequency of analogs to the Local Group in $\Lambda$CDM as $0.09 \times 0.05 = 0.45\ \mathrm{per cent}$. This is hardly reassuring for the $\Lambda$CDM model.
Furthermore, the claim that the look-elsewhere effect needs to be considered does not directly apply to the VPOS around the MW. This is because it was identified, and is typically compared to simulations, by using all 11 classical satellites which were selected by their luminosity, not by searching for the most significant sub-sample. Implementing a look-elsewhere analysis that allows for sub-samples of the 11 classical satellites to be identified as a satellite plane thus allows choices in the simulated sample that were never considered for the observed situation. It therefore over-corrects for an effect that is not present in the observed sample, artificially boosting the degree of agreement with $\Lambda$CDM.

Most recently, Centaurus\,A joined the MW and M31 as a third system for which not only a planar satellite distribution was found, but also an associated kinematic signal in the line-of-sight velocities. A comparison to cosmological simulations was performed that requires only a minimum of parameters of the CASP to be matched simulataneously:\cite{Mueller2018} the on-sky flattening (such that distance uncertainties are of no effect), and the kinematic alignment along the long axis of the on-sky distribution. This avoids the need to address a look-elsewhere effect. The study finds that analog systems to the Centaurus\,A satellite system are rare both in the dark-matter-only Millennium-II simulation (0.1\,per cent), as well as in the hydrodynamical Illustris simulation (0.5\,per cent).

Planes of satellite galaxies have been discovered around at least three systems: the MW and M31 in the Local Group, as well as Centaurus\,A. Analogs of these structures that simulataneously reproduce their spatial (thickness and extend or axis ratio flatenning) and kinematic signatures (line-of-sight velocity correlations or alignment of orbital planes and directions) have been searched in cosmological simulations. Figure \ref{fig2} summarizes some of the comparisons between the observed planes of satellite galaxies and cosmological simulations. For all three systems, such analogs to the observed planes of satellites are very rare in $\Lambda$CDM; they typically occur with a frequency of only 0.1 to 0.5 per cent. There thus should be a similarly extreme satellite plane as one of the observed structures around one out of every $\geq 200$\ satellite systems. Yet three such structures were found around three of the nearest and best-studied host galaxies. This is the Planes of Satellite Galaxies Problem of $\Lambda$CDM. Unless the culprit for this mismatch lies in the details of how cosmological simulations model structure formation, the existence of the observed planes of satellite galaxies poses a serious problem for $\Lambda$CDM cosmology -- and for any competing dark matter model in which satellite galaxies are distributed and orbit similarly (e.g. warm and self-interacting dark matter).

\section{Suggested solutions and open questions}
\label{sect:solutions}

The rarity of satellite arrangements in cosmological simulations that are comparable to the observed planes of satellite galaxies poses a serious problem for the $\Lambda$CDM model. Consequently, numerous possible solutions have been suggested, which have in common that they aim to result in increased phase-space coherence among satellite galaxy systems. They are either based on assuming that the observed satellite galaxies are primordial, or that they are second-generation objects formed in a common event. The former implies that the satellites are of cosmological origin, are dark matter sub-halos which for some reason have a higher degree of phase-space correlation than found in typical cosmological simulations. The latter proposes that the satellites were formed as one population by a process that caused them to share similar orbits. None of the proposed solutions has been generally accepted. This is largely because they all display problems of their own, either due to inconsistencies with observed features, or because major open questions have not yet been conclusively investigated or lead to internal theoretical inconsistencies.

\subsection{Accretion along filaments of the cosmic web}
\label{sect:filaments}

\begin{figure}[ph]
\centerline{\includegraphics[width=5.0in]{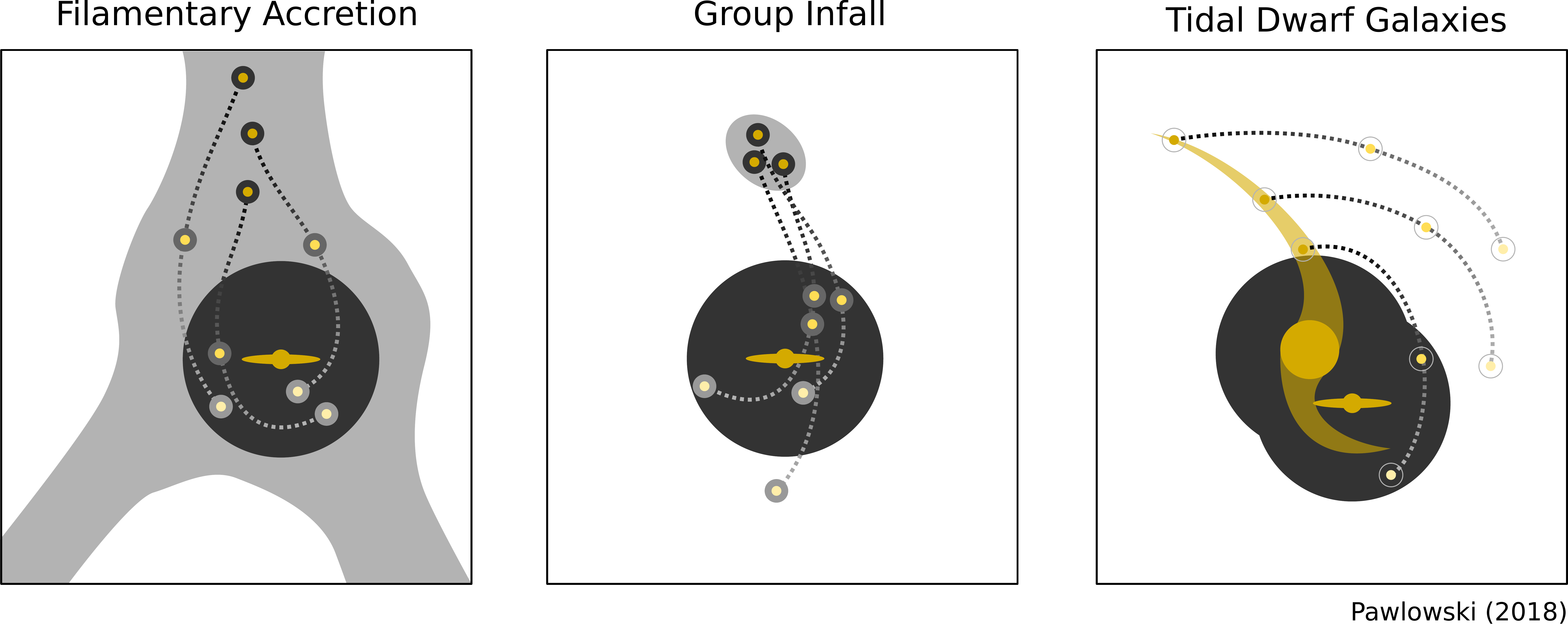}}
\caption{
Illustrations of the three major scenarios that were suggested to explain the correlation of satellite galaxies in rotating planes around a central galaxy (yellow, edge-on disk). Left: the preferential accretion of dwarf galaxies along a small number of cosmic filaments onto the halo of a central host galaxy (Sect. \ref{sect:filaments}); Middle: the accretion of dwarf galaxies in a group (Sect. \ref{sect:groupinfall}); Right: the formation of second-generation Tidal Dwarf Galaxies out of material in the tidal tails of interacting galaxies (Sect. \ref{sect:TDGs}). Three subsequent positions for each of the three sketched dwarf galaxies are shown, connected by dotted lines. The dwarf galaxies in the first two scenarios are embedded in dark-matter (sub-)halos, whereas TDGs are expected to be dark matter free.
 \protect\label{fig3}
 }
\end{figure}

In the hierarchical structure formation of $\Lambda$CDM, matter first collapses along one axis forming a sheet, then along a second axis forming a filament, and then along the filament onto halos residing within, or at nodes connecting several, filaments. This process results in an anisotropic pattern of satellite infall onto host halos, in particular the most massive ones are preferentially accreted along the spine of a filament (left panel in Fig. \ref{fig3}). Concequently, filamentary accretion has been invoked as an explanation for the highly anisotropic planes of satellite galaxies.\cite{Libeskind2005,Zentner2005,Lovell2011,Libeskind2011}

For the planes of satellite galaxies around M31 and Centaurus A -- though not for the MW VPOS -- an alignment with the axis of compression derived from the velocity shear field of observed galaxies in the local Universe has been reported.\cite{Libeskind2015} This orientation is consistent with the preferred direction along which satellite galaxies are channeled towards their hosts in cosmological simulations, and could thus be seen as supporting this explanation for the planes of satellite galaxies. However, in cosmological simulations the accretion of satellites along this direction is boosted by only up to a factor of two compared to uniform accretion,\cite{Libeskind2014} such that it is very doubtful whether the observed satellite planes which consist of at least half of the satellite populations were formed solely by such accretion. Thus, while the accretion along filaments does result in anisotropy in the spatial distribution of satellite galaxies, the effect is not sufficiently strong to explain the highly flattened satellite distributions. Concerning the kinematic correlation of planes of satellite galaxies, it has been argued that filamentary accretion results in a preferred orbital direction of sub-halos in the Aquarius simulations, though no quantitative comparison to the VPOS was provided.\cite{Lovell2011} A subsequent re-analysis of the distribution of orbital poles for satellites in these simulations demonstrated that an as narrow alignment as observed for the orbital poles of the MW satellite galaxies occurs in only 0.6 per cent of realisations, compared to 0.1 per cent for an isotropic distribution.\cite{Pawlowski2012b,Pawlowski2013b} 

One reason why filaments do not produce sufficiently narrow structures is related to their size relative to the host halo onto which satellite are accreted. At the current time in the evolution of the cosmos, filaments feeding halos of virial mass $M_{\mathrm{vir} }\approx 10^{12}\,\mathrm{M}_\odot$, such as the MW and M31, are wider than the hosts' virial radii.\cite{VeraCiro2011} The resulting distribution of material accreted along such wide filaments can therefore not be expected to result in narrow satellite structures of only 20 to 30\,kpc height. While it has been argued that massive sub-halo satellites are accreted preferentially along the central spines of filaments which would give them a more narrow distribution,\cite{Libeskind2005} the effect of preferential accretion along filaments is already self-consistently included in cosmological simulations. Since planes of satellites as extreme as the observed ones are very rare in such simulations, it appears that even such a preferred trajectory of the more massive satellites is not sufficiently strong to explain the observed structures. Furthermore, it is unclear how the preferentially radial accretion along filaments would provide the satellite galaxies with their high, and similarly oriented, angular momentum, in particular if more than one or two filaments contribute to the accretion, or if their orientation changes over time.\cite{Pawlowski2014}

\subsection{Infall of satellite galaxies in groups}
\label{sect:groupinfall}

Another suggested origin of the coherence of satellite galaxy positions and the similarity of their orbital directions is the accretion as members of a common group (middle panel in Fig. \ref{fig3}). If satellites are confined to a group, then upon accretion of this group onto a host galaxy they would share similar orbital orientation, energy, and specific angular momentum.\cite{Lynden-Bell1995} After the initial accretion, tidal forces by the host will disrupt the initially bound group of satellites, which will then disperse along their common orbital plane. To result in the formation of a sufficiently narrow satellite plane (heights of 15 to 30\,kpc), the infalling groups would have to be compact. This allows to test the validity of the group infall hypotesis by determining whether observed dwarf galaxy associations are sufficiently compact. This is not the case. Dwarf galaxy associations in the local Universe are much more spatially extended (150 to 210\,kpc) than the observed planes of satellite galaxies and thus do not support the group infall hypothesis.\cite{Metz2009} 

Nevertheless, motivated by the tight alingment of the LMC and the Magellanic Stream with the VPOS in position and orbital orientation, it has been suggested that this plane of satellites was build by the infall of a group of dwarf galaxies dominated by the LMC.\cite{Donghia2008} The idea that the LMC has brought along a number of satellite galaxies of its own has been further fueled by discoveries of dwarf satellite galaxies in its vicinity. While it is well possible, if not even expected, that massive satellites like the LMC were accompanied by a number of satellite galaxies of their own, their role in the formation of planes of satellite galaxies is questionable. 
Numerical simulations of the accretion of such Magellanic Systems onto the MW have found this scenario to be untenable, because several of the observed VPOS member satellites are situated in regions that are unaccessible to LMC-satellites in the simulations, even if the LMC had completed more than one orbit around the MW.\cite{Nichols2011} A similar scenario has been investigated for the GPoA, in which the dwarf elliptical galaxy NGC\,205, one of the most massive satellites of M31, is assumed to have been the most massive galaxy in an infalling group of dwarf galaxies.\cite{Angus2016} The study found that a GPoA as narrow as observed can be reproduced by their scenario in about 1 per cent of all cases, yet simulataneously reproducing the observed velocities reduces this to 0.1 per cent. The success rate of this scenario is thus comparable to the frequency of arrangements as extreme as the GPoA in cosmological simulations, such that this scenario does not fare better than postulating that the GPoA is a rare outlier. Interestingly, the two satellite systems in the Millennium-II simulation that resemble the GPoA parameters contain very massive satellites that are on their first infall and carry with them an entourage of smaller galaxies.\cite{Ibata2014a}

In cosmological simulations of hosts comparable to the MW and M31, typical groups of sub-halos consist of only 2 to 5 objects,\cite{Li2008,Wetzel2015} and typically less than half of the top 11 satellites are found to have been accreted in groups of two or more objects.\cite{Wang2013,Shao2017}
Thus, it would have to be argued that most of the infalling groups are dark, because only of order one luminous group should have been accreted to form a pronounced, coherently orbiting plane of satellite galaxies. Another related concern is that, to populate the VPOS or GPoA which each consist of at least a dozen satellite galaxies, the progenitor groups would have to contain as many objects. This implies that a substantial fraction of the observed, bright ($M_{\mathrm{V}} \leq -8$) satellite galaxy populations of the MW and M31 would have been accreted in one group, raising the question how a lower-mass group can bring in a comparable number of satellite galaxies as present around the host in the first place. One can attempt to address this issue by postulating that not only one but several group accretions contributed to the formation of a satellite galaxy plane, but is then left to find an explanation why several groups were accreted into the same orbital plane, in the same orbital direction, and at about the same time.

Ultimately, like the accretion along filaments, group infall is already self-consistently included in cosmological simulations. While some of the anisotropy in satellite systems can be attributed to this effect, it is apparently not sufficiently strong to result in a high frequency of analogs to the observed planes of satellite galaxies. Saving this suggested solution would require a process that enhances the group infall signal, such as by populating only select groups with large numbers of luminous satellite galaxies, possibly due to environmental conditions during their formation. No candidate for such a process has been proposed yet.

\subsection{Hydrodynamics and baryonic physics}
\label{sect:hydro}

The inclusion of baryonic physics in cosmological simulations helps to alleviate several small-scale problems of $\Lambda$CDM, such as the Missing Satellites and the Core-Cusp Problem.\cite{Bullock2017} It is therefore maybe an understandable initial reaction to the Planes of Satellite Galaxies Problem to invoke the same argumentative structure. The problem was first identified by comparisons with collisionless, dark-matter only simulations, but what we observe are luminous galaxies formed through the complex interplay of gas physics, star formation, and feedback processes. So maybe the inclusion of realistic baryonic physics in cosmological simulations can resolve the Planes of Satellite Galaxies Problem? Unfortunately, this particular problem is not that easily addressed. While other small-scale problems are related to the internal structure of satellite galaxies, the Planes of Satellite Galaxies Problem is concerned with their distribution and motion in their host halo, on scales of 100s of kpc. Baryonic processes acting in the central regions of dark matter (sub-)halos do not have an immediate effect on the positions or orbital motions of the halos, and thus it is a priori implausible that baryonic physics has any effect on the emergence of coherences in the phase-space distribution of satellite galaxies. The most significant difference between dark-matter-only and hydrodynamical simulations is rather an indirect one: the existence of a central host galaxy changes the potential in a way that increased the tidal disruption of satellites in the inner region of a host halo. This results in a less radially concentrated satellite distribution.\cite{GarrisonKimmel2017}

Thus far, only very few comparisons to hydrodynamic simulations have been published. This is largely due to lacking sample size, because running such simulations with high enough resolution to resolve a sufficient number of satellite galaxies is computationally expensive. Focussing on the VPOS, one study claimed that the Planes of Satellite Galaxies Problem is solved by hydrodynamical cosmological simulations,\cite{Sawala2014} but did not provide a quantitative comparison, nor a comparison with existing dark-matter-only analogs of the analysed simulations. A re-analysis of the claims, including a comparison to dark-matter-only simulations, revealed that there is no evidence that the modelling of baryonic processes has improved the match between the observed satellite planes and the simulated satellite systems. Both the plane flattening as well as the orbital alignment of satellites in their simulation was found to be consistent with that typical for dark-matter-only simulations.\cite{Pawlowski2015b} 

In a more detailed comparison between four hydrodynamical zoom simulations of satellite systems and their dark-matter only equivalents, motivated by the GPoA, it was found that planes fitted to satellites in the hydrodynamical simulations are more statistically significant compared to an isotropic situation.\cite{Ahmed2017} This is due to the larger radial extent of satellite systems in those simulations, which reduces the probability of finding a given plane height if the angular positions of satellite are randomized. However, none of the simulated systems contained a satellite plane analog as extreme as the observed GPoA.

Analogs of the Centaurus\,A plane are also similarly rare in the dark-matter-only Millennium-II (0.1 per cent) and the hydrodynamical Illustris simulation (0.5 per cent), with the difference not being statistically significant.\cite{Mueller2018}

In summary, there is currently neither evidence that baryonic physics enhances the frequency of analogs of the observed planes of satellite galaxies in cosmological simulations, nor is it obvious what baryonic physics process would increase the coherence in the phase-space distribution of dark-mater-dominated satellite galaxy systems.

\subsection{Tidal Dwarf Galaxies}
\label{sect:TDGs}

Tidal Dwarf galaxies (TDGs) are second-generation galaxies formed when collisional debries around interacting galaxies collapse under self-gravity and form objects consisting of stars and gas that have properties (sizes, masses, star formation rates) similar to dwarf galaxies.\cite{Weilbacher2000,BournaudDuc2006,Wetzstein2007} TDGs can survive their internal feedback processes as well as external tidal disruption for at least several Gyr.\cite{Ploeckinger2015} Regarding planes of satellite galaxies, TDGs offer an intriguing explanation because they form out of a common tidal tail, and therefore share the same orbital plane and direction (right panel in Fig. \ref{fig3}).\cite{Kroupa2005,Metz2007b,Pawlowski2012a,Kroupa2012a,Hammer2013} The phase-space distribution of tidal debris can therefore easily explain satellite galaxies co-orbiting in one plane, and can even account for a counter-orbiting population in both galaxy merger and fly-by interactions.\cite{Pawlowski2011} The formation of TDGs has been observed, and they thus should form in any realistic cosmological model. Indications for their formation have indeed been found in the $\Lambda$CDM hydrodynamical simulation EAGLE.\cite{Ploeckinger2017}

Historically, a progenitor to the TDG hypothesis was the first attempt to explain the alignment of the MW satellite galaxies along a great circle. It was proposed that a Greater Magellanic Galaxy, during a past encounter with the MW, fragmented into several smaller galaxies that then spread along similar orbits, with the LMC and SMC constituting the largest fragments.\cite{LyndenBell1976,Kunkel1976} While this particular scenario is untenable if the LMC is on its first infall onto the MW,\cite{Besla2007} it illustrates that the TDG hypothesis requires a past galaxy interaction and can thus be ruled out if no viable interaction scenario exists. In the Local Group, three scenarios for past interactions are being discussed: a past merger of a galaxy with the MW,\cite{Pawlowski2012a} a fly-by interaction between the MW and M31,\cite{Zhao2013,Bilek2017} and a past merger forming M31.\cite{Hammer2010} The latter has been studied in detail via numerical simulations, which indicate that a polar merger with 3:1 mass ratio starting about 9\,Gyr ago is consistent with current properties of M31 (disk, bulge, streams, and distribution of stellar ages), and could also have produced TDGs in an orbital configuration that coincides with the GPoA.\cite{Hammer2013} It has even been argued that TDGs were expelled from this encounter towards the MW, where they can help form the VPOS.\cite{Fouquet2012}

While the TDG hypothesis is very successful in explaining strong phase-space correlations, several criticisms have been raised. One concerns the low metallicity of the satellite dwarf galaxies. TDGs form out of the debris of more massive galaxies which are more efficient in holding onto their gas as they evolve. Their material should therefore be pre-enriched with metals. Consequently, TDGs are expected to have a higher metallicity than primordial dwarf galaxies of the same stellar mass or luminosity. If the observed dSphs forming satellite planes are TDGs, they should then be offset from the well established mass-metallicity relation for dwarf galaxies.\cite{Kirby2013} Such an offset is not seen in the data, and furthermore the on- and off-plane satellite galaxies around M31 do not show differences in other observable properties (sizes, luminosities, masses, and star formation histories).\cite{Collins2015} However, it has been argued that if the satellite planes were formed out of TDGs, this event would have had to have happened very early in the Universe (redshift $z \approx 2$, or about 10\,Gyr ago).\cite{Pawlowski2012a} At that time, larger-mass galaxies were less pre-enriched, which can help alleviate this concern.\cite{Recchi2015}

The maybe most fundamental issue with the TDG scenario is the dark matter content of these galaxies. The observed dSphs in the Local Group display high velocity dispersions for their luminosities, resulting in mass-to-light ratios exceeding $M/L = 10\,\mathrm{M}_\odot/\mathrm{L}_\odot$.\cite{McConnachie2012} Consequently, in the $\Lambda$CDM framework they are interpreted as being highly dark matter dominated. TDGs, in contrast, should be devoid of dark matter. This is because the galactic disk of their parent galaxy is rotation-supported and dynamically cold and can thus form tidal tails, while the parent's dark matter halo is pressure-supported and dynamically too hot to contribute to the formation of narrow tidal tails. To save the TDG hypothesis, the apparent dark matter content of observed dSphs can be addressed in one of three ways: (1) question the validity of the observed high velocity dispersions or the assumption that the observed dSphs are in dynamical equilibrium,\cite{Kroupa1997,Metz2007b,Yang2014} (2) modify the properties of dark matter such that it can participate in TDG formation,\cite{Foot2013,Randall2015} or (3) change from a dark matter to a modified gravity framework in which all objects behave as if they had dark matter if falsely interpreted with Newtonian dynamics.

The last choice is maybe the most extreme, but also the most promising one. It implies that the $\Lambda$CDM model can be abandoned in favor of a modified gravity model, such as one based on Modified Newtonian Dynamics (MOND).\cite{Milgrom1983,FamaeyMcGaugh2012} If the observed dark matter signatures in galaxies are interpreted as a sign of different gravitational dynamics, then all objects experiencing sufficiently low internal and total accelerations should show higher accelerations than expected under the assumption of Newtonian dynamics. In other words, in a modified gravity universe interpreted under the false assumption of Newtonian dynamics, all objects would look like they are dark matter dominated. This is particularly true for TDGs, and supported by the finding that there appears to be only one population of dwarf galaxies in the Universe,\cite{Dabringhausen2013} whereas in $\Lambda$CDM and related models the so-called Dual Dwarf Galaxy Theorem\cite{Kroupa2012a} should be true, according to which both a dark matter dominate population of primordial dwarf galaxies as well as a dark matter free population of TDGs should exist (though this depends on the relative abundances and survival timescales of these galaxies). Numerical simulations of galaxy collisions have demonstrated that TDGs can form in MOND,\cite{TiretCombes2008,Renaud2016} and their formation is potentially even enhanced due to the higher self-gravity of collapsing gas in the tidal tails compared to the Newtonian case. Observationally, there is currently no concensus on whether young TDGs display apparent mass discrepancies and flat rotation curves that are consistent with MOND\cite{Bournaud2007}, follow Newtonian dynamics\cite{Lelli2015}, or whether they are even in sufficient equilibrium to allow such tests while they are still actively forming.\cite{Flores2016}

Due to the very different scenarios, and different expectations for the dynamics and abundance of TDGs, it is difficult to conclusively confirm or refute this hypothesis. One strong prediction of the TDG hypothesis is that host galaxies which never experienced a major galaxy encounter should not host planes of satellite galaxies (no TDGs have formed around them). If planes of satellites were found to be common around isolated host galaxies with a quiet merger history, the TDG hypothesis would have to be abandoned as a viable origin of planes of satellite galaxies in any cosmological model.

\section{Summary and outlook}
\label{sect:outlook}

Flattened structures of satellite galaxies with corresponding kinematic correlations that are indicative of co-orbiting satellites have been found around the closest and best-studied satellite galaxy systems: the Vast Polar Structure (VPOS) around the Milky Way, the Great Plane of Andromeda (GPoA), and the Centaurus A Satellite Plane (CASP). In cosmological simulations based on the currently favored $\Lambda$CDM model of cosmology, satellite distributions that are similarly flattened and kinematically correlated are extremely rare, with some studies reporting expected frequencies on the order of 0.1 per cent for each of the three. The existence of the VPOS, GPoA, and CASP thus constitutes a severe {\it Planes of Satellite Galaxies Problem} for the standard model of cosmology.

While processes that cause correlations among satellite galaxy systems -- preferential accretion of dwarf galaxies along filaments and in groups -- have been identified in cosmological simulations, the strength of the resulting overall anisotropy is insufficient to explain the observed planes of satellite galaxies. The inclusion of baryonic physics in cosmological simulations offers no obvious enhancement of the phase-space correlations among dark-matter dominated satellite galaxies. The formation of tidal dwarf galaxies (TDGs) can result in highly phase-space correlated populations of dwarf galaxies, but their expected lack of dark matter makes these an often disregarded solution to the Planes of Satellite Galaxies Problem, unless one is willing to consider a paradigm change to alternative cosmological models based on modified gravity or more exotic types of dark matter.

A full understanding of the origin of the Planes of Satellite Galaxies Problem will require both theoretical developments and observational advances. Studying the dynamics and survival timescales of all observed planes of satellite galaxies, as well as implications for the host galaxy's potential, is a logical next step that has already been taken for the GPoA.\cite{Fernando2017} The ever increasing resolution and detail with which physical processes are modelled in simulations will improve the statistics of comparisons between the observed structures and model expectations to further constrain the degree of tension with $\Lambda$CDM, and support investigations of environmental effects\cite{PawlowskiMcGaugh2014b} and correlations with host halo properties.\cite{Buck2015} Taken together with other small-scale enigmas that challenge the $\Lambda$CDM model, the Planes of Satellite Galaxies Problem can also encorage the study of alternative models such as MOND and hybrid approaches that aim to combine the successes of the dark matter and modified gravity hypotheses.\cite{Blanchet2008,Berezhiani2015}

Observationally, first attempts to statistically quantify the prevalence of planes of satellite galaxies in the Universe have been made.\cite{Ibata2014,Ibata2015} However, these are hampered by two issues: the intrinsic difficulty of studying the full 6D phase-space distribution of dwarf satellite galaxies beyond the Local Group, and the small number of observed satellites per host. Distance measurements are often uncertain to at least 5 to 10 per cent, which exceeds the extent of a MW-like satellite system if the host is at a distance of 5\,Mpc or more, such that only projected positions can be considered. Measurements of dwarf galaxy proper motions are unfeasible beyond the Local Group because the expected angular speeds of distant satellites are orders of magnitude smaller than the best astrometric precision achivable today, such that only line-of-sight velocities are observationally accessible. Finally, for most host galaxies only the one or two brightest satellite galaxies have been observed by surveys such as the SDSS, which prohibits a detailed analysis of mutual phase-space correlations between satellites.

Due to these unavoidable difficulties in expanding studies of phase-space correlations to more distant hosts, it is crucial to obtain at least the set of achievable observables: the projected positions and line-of-sight velocities of as many satellite galaxies per host as possible. Some initatives that hold potential to provide this data for a larger sample of systems are already underway. Two examples are the SAGA survey which spectroscopically identifies satellite galaxy populations around analogs of the Milky Way at distances of 20 to 40\,Mpc, \cite{Geha2017} and the DGSAT survey which employs amateur telescopes to identify low-surface-brightness satellite galaxies around more nearby hosts.\cite{Javanmardi2016}. Future surveys and facilities such as extremely large telescopes will make it feasible to study the phase-space structure of systems of satellite galaxies, and correlations therein, around even more distant hosts.

\section*{Acknowledgments}

Support for this work was provided by NASA through Hubble Fellowship grant \#HST-HF2-51379.001-A awarded by the Space Telescope Science Institute, which is operated by the Association of Universities  for  Research  in  Astronomy,  Inc.,  for  NASA,  under  contract  NAS5-26555

\bibliographystyle{ws-mpla}
\bibliography{review}

\end{document}